Reply to Mills et al.: Oceanic Anoxic Event, a mechanism for selecting animals with the ability to survive hypoxic conditions.

Saint-Antonin, Francois, Univ. Grenoble Alpes, CEA, LITEN, L2N, 17 rue des Martyrs, F-38054 Grenoble Cedex 9, France.

francois.saint-antonin@cea.fr

Abstract

It is generally considered that animal life was triggered by the rise of oxygen levels. Based on experiments evaluating the minimum range of oxygen levels at which sponges can survive, Mills et al. [1] defend the opposite view. However, the authors do not demonstrate that 'animal life was not triggered by the oxygen rise' is the only possible and unique conclusion from their observation. In this reply, it is suggested that a mechanism to explain the ability of sponges to survive at low oxygen biota is 'Ocean Anoxic Events'. These lead to oxygen depletion and a series of them would selectively favor animals able to survive at low oxygen levels. Thus, the origin of the ability of marine animals to survive in low oxygen biota remains to be clarified.

It is generally considered that "low levels of atmospheric oxygenation delayed the origin of animals until 850-542 million years" [1]. By 580 Ma, there was a general oxygenation of the oceans "likely requiring a minimum of 10% of present atmospheric levels" [1]. By that time, metazoan (Ediacara macrobiota) had appeared with "relatively high oxygen demands compared with the aerobic microbes that thrived earlier in the Proterozoic Eon" [1]. Combining these observations, it was concluded that animal life was triggered by the rise of oxygen levels. Based on experiments evaluating the minimum range of oxygen levels at which sponges can survive (i.e. 0.5-4.0%), Mills et al. [1] defend the opposite view, namely that animal life was not triggered by the oxygen rise. Sponges were chosen as being "the last common ancestor of metazoans likely to exhibit a physiology and morphology similar to that of a modern sponge". The oxygen range of 0.5-4% is very low compared to present day oxygen level ($\approx$20%) and can be considered as a hypoxic state.

However, the authors do not demonstrate that 'animal life was not triggered by the oxygen rise' is the only possible and unique conclusion from their observation and, other possibilities concerning the ability of sponges to survive in low oxygen range biota are not explored or considered by them.

Here, I would like to propose a different possible origin for the sponge's ability to survive in low oxygen biota. This is related to Oceanic Anoxic Events (OAE). An OAE occurs when there is a large depletion in oxygen below ocean surface levels in a relatively 'rapid way'. Geologists have extensively described and documented several tens of OAE during the Paleozoic and Mesozoic periods. In some cases, OAE were themselves the consequence of another catastrophic event, such as a large volcanic eruption leading to large igneous province also inducing substantial oxygen depletion. Many OAE were the direct cause of several large marine mass extinctions at Earth scale. By favoring marine animals with the ability to survive hypoxic-anoxic conditions and killing the others, the oxygen depletion would have had a selective impact. Thus, repeated OAE must have eradicated marine animals dependent only on high oxygen levels. Returning to the ability of sponges to survive in low oxygen biota, the question is whether this might simply be the selective result of successive OAE over several hundred million years ?